\documentclass[twoside]{article}
\usepackage{qic}

\usepackage{amsmath}
\usepackage{amssymb}
\usepackage{braket}

\newcommand{\ourtitle}{Limit theorems for quantum walks with memory}

\begin{document}
\setlength{\textheight}{8.0truein}    %FOR 2ND PAGE ONWARDS

\runninghead{\ourtitle}
            {N. Konno and T. Machida}

\normalsize\textlineskip
\thispagestyle{empty}
\setcounter{page}{1}

%\copyrightheading{Vol.}{No.}{Year}{Page Nos.}
%\copyrightheading{0}{0}{2003}{000--000}

\vspace*{0.88truein}

\alphfootnote

\fpage{1}

\centerline{\bf
%%%%%%%%%%%%%%%%%%%%%
%Put in titiles here
%%%%%%%%%%%%%%%%%%%%%
\ourtitle}
\vspace*{0.37truein}
\centerline{\footnotesize
%%%%%%%%%%%%%%%%%%%%%%%%%%%%%%%%%%%%
%put authors' name and address here
%%%%%%%%%%%%%%%%%%%%%%%%%%%%%%%%%%%%
Norio Konno \, and \, Takuya Machida}
\vspace*{0.015truein}
\centerline{\footnotesize\it Department of Applied Mathematics, Faculty of Engineering,}
\baselineskip=10pt
\centerline{\footnotesize\it Yokohama National University, Hodogaya, Yokohama, 240-8501, Japan}
\vspace*{0.225truein}
%\publisher{(received date)}{(revised date)}

\vspace*{0.21truein}

%% \abstracts{first paragraph}{second paragraph}{third paragraph}
%% If there is only one paragraph, just keep the second and third empty 
%% like the following one 
\abstracts{
%%%%%%%%%%%%%%%%%%%%
% put abstract here
%%%%%%%%%%%%%%%%%%%%
Recently Mc Gettrick \cite{gettrick} introduced and studied a discrete-time 2-state quantum walk (QW) with a memory in one dimension. He gave an expression for the amplitude of the QW by path counting method. Moreover he showed that the return probability of the walk is more than 1/2 for any even time. In this paper, we compute the stationary distribution by considering the walk as a 4-state QW without memory. Our result is consistent with his claim. In addition, we obtain the weak limit theorem of the rescaled QW. This behavior is strikingly different from the corresponding classical random walk and the usual 2-state QW without memory as his numerical simulations suggested.
}{}{}

\vspace*{10pt}

\keywords{4-state quantum walk, quantum walk with memory, localization}
\vspace*{3pt}
%\communicate{to be filled by the Editorial}

\vspace*{1pt}\textlineskip    %) USE THIS MEASUREMENT WHEN THERE IS
   %) A SECTION HEADING
%\vspace*{-0.5pt}
%\noindent

\bibliographystyle{qic}

\section{Introduction}
The quantum walk (QW) is a counterpart of the classical random walk. The QW has two types like the random walk, that is, one is discrete-time and the other is continuous-time. In this paper we focus on the discrete-time case. Let $\mathbb{Z}$ be the set of integers. The 2-state QW on $\mathbb{Z}$ has been well studied by various authors \cite{ambainis_2001,kempe,kendon,konno_2008_2}. For example, the weak limit theorem was given in \cite{konno_2002_1,konno_2005_1}. 

Mc Gettrick \cite{gettrick} introduced and investigated 2-state QWs with one-step memory (or also called ``with 2nd order") on $\mathbb{Z}$. We consider the walk as a 4-state QW without memory by rewriting the state $\ket{n_2,n_1,p}(=\ket{n_2}\otimes\ket{n_1}\otimes\ket{p})$ with $(n_2,n_1,p) \in \mathbb{Z}^2 \times \{0,1\}$ in his model as $\ket{n_1,n_1 - n_2 + 1 +p}(=\ket{n_1}\otimes\ket{n_1-n_2+1+p})$ with $(n_1,n_1-n_2+1+p) \in \mathbb{Z} \times \{0,1,2,3\}$ in our 4-state model. The integers $n_1, \> n_2 \in \mathbb{Z}$ correspond to the position and $p \in \{0,1\}, \> n_1 - n_2 + 1 +p \in \left\{0,1,2,3\right\}$ mean the (coin) state or the chirality. He derived an expression for the amplitude of his QW by path counting method. Let $\mathbb{P}(X_t=x)$ be the probability that the quantum walker, $X_t$, exists at position $x\in\mathbb{Z}$ at time $t$ starting from the origin. He claimed that $\mathbb{P}(X_{2t}=0)\geq 1/2$ for any $t$ by induction on the time step. Corresponding to his result, we calculate the stationary distribution of $X_t$ and see that the return probability $\mathbb{P}(X_{2t}=0)$ converges to $2-\sqrt{2}=0.58578\cdots$, as $t\rightarrow\infty$, by using the Fourier analysis. The result is consistent with his claim. In general, if $\limsup_{t\rightarrow\infty}\mathbb{P}(X_t=0)>0$, we say that localization occurs. Therefore our result insists that localization occurs for an initial state. Related with quantum physics, localization of the QW has been investigated \cite{banuls,chisaki,inui1,inui3,konno_2009_2,wojcik}. Moreover he numerically reported that the behaviour of the 2-state QW with one-step memory, equivalently the 4-state QW without memory, is remarkably different from the corresponding classical random walk and the ordinary 2-state QW without memory. By the Fourier analysis again, we show that $X_t/t$ converges weakly to a random variable as $t\rightarrow\infty$. The limit measure is described by both a $\delta$-function corresponding to localization and a density function. A similar limit theorem was presented in Inui et al. \cite{inui1} for the 3-state Grover walk. 
4-state models were also studied in Brun et al. \cite{brun}, Venegas-Andraca et al. \cite{venegas2005} and Segawa and Konno \cite{Segawa}.
Their 4-state walks have a 3-direction shift operator (i.e., left, right, and center).
On the other hand, the walker in our 4-state walk moves to the left or right.

The rest of the present paper is organized as follows. Section 2 treats the definition of our 4-state QW. In Section 3, we present the limit theorems as our main result. Section 4 is devoted to proofs of the theorems. Summary is given in the last section.

%%%%%%%%%%%%%%%%%%%%%%%%%%%%%%%%%%%%%%%%%%%%%%%%%%%%%%
\section{Definition of 4-state QW}

In this section we define a 4-state QW without memory corresponding to the 2-state QW with one-step memory introduced and studied by Mc Gettrick \cite{gettrick}. In the paper, the time evolution for his QW (called case (c) there) is written by the following operators:
\begin{eqnarray}
 C_1 &:& \begin{array}{l}
  \ket{n+1,n,0}\longrightarrow a\ket{n+1,n,0} + b\ket{n+1,n,1},\\
	   \ket{n+1,n,1}\longrightarrow c\ket{n+1,n,0} + d\ket{n+1,n,1},\\
	   \ket{n-1,n,0}\longrightarrow a\ket{n-1,n,0} + b\ket{n-1,n,1},\\
	   \ket{n-1,n,1}\longrightarrow c\ket{n-1,n,0} + d\ket{n-1,n,1},
	  \end{array}\label{eq:coin}\\
 C_2 &:& \begin{array}{l}
	\ket{n+1,n,0}\longrightarrow \ket{n-1,n,0},\\
	       \ket{n+1,n,1}\longrightarrow \ket{n+1,n,1},\\
	       \ket{n-1,n,0}\longrightarrow \ket{n+1,n,0},\\
	       \ket{n-1,n,1}\longrightarrow \ket{n-1,n,1},
	      \end{array}\nonumber\\
 S &:& \begin{array}{l}
  \ket{n+1,n,0}\longrightarrow \ket{n,n-1,0},\\
	     \ket{n+1,n,1}\longrightarrow \ket{n,n-1,1},\\
	     \ket{n-1,n,0}\longrightarrow \ket{n,n+1,0},\\
	     \ket{n-1,n,1}\longrightarrow \ket{n,n+1,1},
	    \end{array}\nonumber
\end{eqnarray}
where $\ket{n_2, n_1, p}$ is the state, $a, b, c, d\,\in\mathbb{C}$ are amplitudes and $\mathbb{C}$ is the set of complex numbers. Here the meaning of each element of $\ket{n_2, n_1, p}$ is as follows;  $n_2 \in\mathbb{Z}$ is the previous position (corresponding to the one-step memory), $n_1 \in\mathbb{Z}$ is the current position, and $p \in \{0,1\}$ is the coin state (or chilarilty). The state space of his 2-state QW with one-step memory is composed of the set of the following vectors: 
\begin{eqnarray*}
\ket{n-1,n,0}, \quad \ket{n-1,n,1}, \quad \ket{n+1,n,0}, \quad \ket{n+1,n,1} \qquad (n \in \mathbb{Z}).
\end{eqnarray*}
By rewriting $\ket{n_2,n_1,p}$ in his setting as $\ket{n_1,n_1 - n_2 + 1 +p}$ in our 4-state model, the state space of the 4-state QW is composed of 
\begin{eqnarray*}
\ket{n,0}, \quad \ket{n,1}, \quad \ket{n,2}, \quad \ket{n,3} \qquad (n \in \mathbb{Z}).
\end{eqnarray*}
For the usual 2-state QW (without memory) with left chirality state $\ket{0}(=\ket{L})$ and right chirality state $\ket{1}(=\ket{R})$, we put, for example, 
\begin{eqnarray*}
\ket{0} = {}^T [1,0] \quad \ket{1} = {}^T [0,1],
\end{eqnarray*}
where $T$ is the transposed operator. In a similar fashion, for our 4-state QW  (without memory), we put 
\begin{align*}
\ket{0} &= {}^T [1,0,0,0], 
\quad \ket{1} = {}^T [0,1,0,0], 
\\
\ket{2} &= {}^T [0,0,1,0], 
\quad \ket{3} = {}^T [0,0,0,1].
\end{align*}
By using this notation, the operator defined by (\ref{eq:coin}) becomes
\begin{eqnarray}
 C_1 &:& \begin{array}{l}
  \ket{n,0}\longrightarrow a\ket{n,0} + b\ket{n,1},\\
	   \ket{n,1}\longrightarrow c\ket{n,0} + d\ket{n,1},\\
	   \ket{n,2}\longrightarrow a\ket{n,2} + b\ket{n,3},\\
	   \ket{n,3}\longrightarrow c\ket{n,2} + d\ket{n,3},
	  \end{array}\label{eq:coin_2}\\
 C_2 &:& \begin{array}{l}
	\ket{n,0}\longrightarrow \ket{n,2},\\
	       \ket{n,1}\longrightarrow \ket{n,1},\\
	       \ket{n,2}\longrightarrow \ket{n,0},\\
	       \ket{n,3}\longrightarrow \ket{n,3},
	      \end{array}\nonumber\\
 S &:& \begin{array}{l}
  \ket{n,0}\longrightarrow \ket{n-1,0},\\
	     \ket{n,1}\longrightarrow \ket{n-1,1},\\
	     \ket{n,2}\longrightarrow \ket{n+1,2},\\
	     \ket{n,3}\longrightarrow \ket{n+1,3},
	    \end{array}\nonumber
\end{eqnarray}
where $SC_2$ is shift operator $S$ of his model.
The amplitudes of $C_1$ are described in Table \ref{tb:operator_c}. 
\begin{table}[h]
 \begin{center}
   \begin{tabular}{|c||c|c|c|c|}\hline
    &$\ket{n,0}$ & $\ket{n,1}$ & $\ket{n,2}$ & $\ket{n,3}$ \\\hline\hline
    $\ket{n,0}$ & $a$& $c$& 0& 0\\\hline
    $\ket{n,1}$ & $b$& $d$& 0& 0\\\hline
    $\ket{n,2}$ & 0& 0& $a$& $c$\\\hline
    $\ket{n,3}$ & 0& 0& $b$& $d$\\\hline
   \end{tabular}\\[5mm]
   \tcaption{Amplitudes of $C_1$}
  \label{tb:operator_c}
 \end{center}
\end{table}

\begin{figure}[h]
 \begin{center}
  \includegraphics[scale=1]{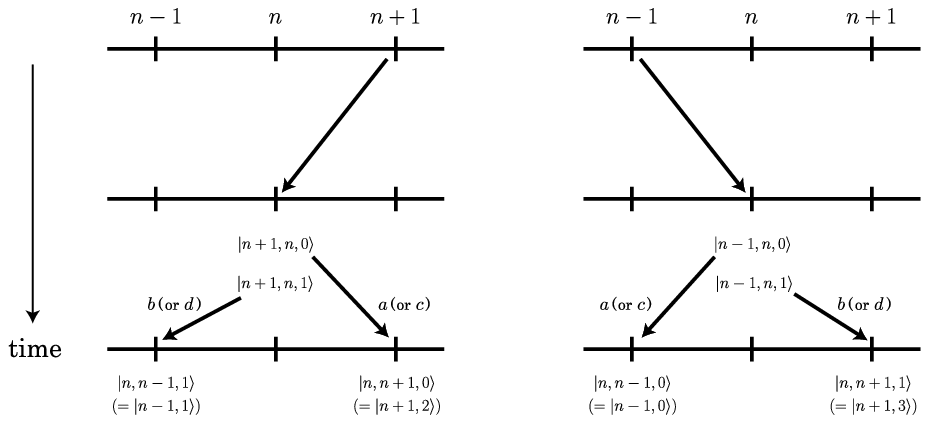}\\[1cm]
  \includegraphics[scale=0.8]{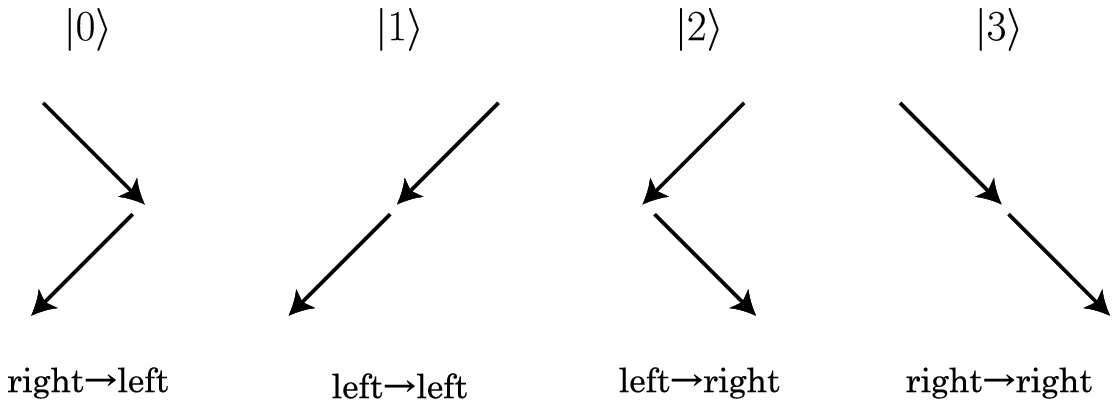}\\[5mm]
  \fcaption{Relation between the states $\ket{0},\ket{1},\ket{2},\ket{3}$ and the walk with memory.}
  \label{fig:relation}
 \end{center}
\end{figure}

The matrix expression of $C_1$ becomes 
\[
 \tilde{C_1}=\left[\begin{array}{cccc}
    a& c& 0& 0\\
	  b & d & 0& 0\\
	  0 & 0 & a & c\\
	  0 & 0 & b &d
	 \end{array}\right].
\]
In a similar way, the permutation matrix $\tilde{C_2}$ determined by $C_2$ is
\[
 \tilde{C_2}=\left[\begin{array}{cccc}
    0& 0& 1& 0\\
	  0 & 1 & 0& 0\\
	  1 & 0 & 0 & 0\\
	  0 & 0 & 0 &1
	 \end{array}\right].
\]
That is, we exchange $\ket{0}$ with $\ket{2}$.
The matrix $\tilde{S}$ corresponding to $S$ is given by the following:
\[
 \tilde{S}=\sum_{n\in\mathbb{Z}}\ket{n-1}\bra{n}\otimes(\ket{0}\bra{0}+\ket{1}\bra{1})+\ket{n+1}\bra{n}\otimes(\ket{2}\bra{2}+\ket{3}\bra{3}).
\]
Therefore, the one-step time evolution operator is given by $\tilde{S}(\tilde{I}\otimes\tilde{C_2}\tilde{C_1})$, where $\tilde{I}$ is the infinite identity matrix.
We should remark that
\[
 \tilde{C_2}\tilde{C_1}=\left[\begin{array}{cccc}
    0&0 &a &c \\b&d&0&0 \\ a&c&0&0 \\ 0&0&b&d
	 \end{array}\right].
\]
The operator $S$ becomes the position shift operator. Then it determines the dynamics of our 4-state QW. In other words, $\ket{0}$ and $\ket{1}$ correspond a left-mover and  $\ket{2}$ and $\ket{3}$ correspond a right-mover.
The states $\ket{0},\ket{1},\ket{2}$ and $\ket{3}$ correspond to the movement ``right$\to$left'', ``left$\to$left'', ``left$\to$right'' and ``right$\to$right'', respectively (See Figure \ref{fig:relation}).

From now on we define the 4-state QWs. Let $\ket{x}$ ($x\in\mathbb{Z}$) be an infinite-component vector which denotes the position of a walker. Here $x$-th component of $\ket{x}$ is 1 and the others are 0.
Let $\ket{\psi_{t}(x)} \in \mathbb{C}^4$ be the amplitude of the walker at position $x$ at time $t$.
The 4-state QW at time $t$ is expressed by
\begin{equation*}
 \ket{\Psi_t}=\sum_{x\in\mathbb{Z}}\ket{x}\otimes\ket{\psi_{t}(x)}\,\left(=\sum_{x\in\mathbb{Z}}\ket{x,\psi_{t}(x)}\right).
\end{equation*}
Rewriting $\ket{n\pm 1,n,p}$ in Mc Gettrick \cite{gettrick} as $\ket{n, \mp 1 + 1 + p}$ in our 4-state model, his time evolution can be described as the following $4\times 4$ unitary matrix:
\begin{equation*}
 U(=\tilde{C_2}\tilde{C_1})=\left[\begin{array}{cccc}
    0&0 &a &c \\b&d&0&0 \\ a&c&0&0 \\ 0&0&b&d
	 \end{array}\right].
\end{equation*}
Noting that $\ket{0}$ and $\ket{1}$ correspond a left-mover and  $\ket{2}$ and $\ket{3}$ correspond a right-mover, $U$ is divided into $P$ and $Q$ as follows:
\begin{equation*}
 P=\left[\begin{array}{cccc}
    0&0 &a &c \\b&d&0&0 \\ 0&0&0&0 \\ 0&0&0&0
	 \end{array}\right],\quad
 Q=\left[\begin{array}{cccc}
    0&0 &0 &0 \\0&0&0&0 \\ a&c&0&0 \\ 0&0&b&d
	 \end{array}\right].
\end{equation*}
Therefore the evolution is determined by
\begin{equation}
 \ket{\psi_{t+1}(x)}=P\ket{\psi_t(x+1)}+Q\ket{\psi_t(x-1)}, \label{eq:te}
\end{equation}
as in the case of usual 2-state QW without memory. This implies that $P$ is a weight of the left movement and $Q$ is the right one. The probability that the quantum walker $X_t$ is at position $x$ at time $t$, $\mathbb{P}(X_t=x)$, is defined by
\begin{equation*}
 \mathbb{P}(X_t=x)=||\ket{\psi_t(x)}||^2,
\end{equation*}
where $||\ket{x}||^2=\braket{x|x}$.

For example, we calculate $\mathbb{P}(X_2=0)$ and $\mathbb{P}(X_4=0)$ under condition $a=b=c=-d=1/\sqrt{2}$ and $\ket{\psi_0(0)}={}^T[\alpha_1,\alpha_2,\alpha_3,\alpha_4],\, \ket{\psi_0(x)}={}^T[0,0,0,0]\,(x\neq 0)$, since Mc Gettrick \cite{gettrick} computed the same quantity.
From (\ref{eq:te}), we have
\begin{align*}
 \ket{\psi_2(0)}\,&=P\ket{\psi_1(1)}+Q\ket{\psi_1(-1)}\\
 &=PP\ket{\psi_0(2)}+PQ\ket{\psi_0(0)}+QP\ket{\psi_0(0)}+QQ\ket{\psi_0(-2)}\\
 &=(PQ+QP)\ket{\psi_0(0)}\\
 &=\frac{1}{2}\left[\begin{array}{cccc}
    1& 1& 1& -1\\ 0&0&0&0\\ 1&-1&1&1\\ 0&0&0&0
	 \end{array}\right]
 \left[\begin{array}{c}
  \alpha_1\\ \alpha_2\\ \alpha_3\\ \alpha_4
       \end{array}\right],\\[5mm]
 \ket{\psi_4(0)}\,&=(PQ^2P+Q^2P^2+QPQP+P^2Q^2+PQPQ+QP^2Q)\ket{\psi_0(0)}\\
 &=\frac{1}{4}\left[\begin{array}{cccc}
    2& 0& 2& 0\\ 1&1&-1&1\\ 2&0&2&0\\ -1&1&1&1
	 \end{array}\right]
 \left[\begin{array}{c}
  \alpha_1\\ \alpha_2\\ \alpha_3\\ \alpha_4
       \end{array}\right].
\end{align*}
If $\alpha_1=\alpha_2=\alpha_3=\alpha_4=1/2$, we have
\begin{equation}
 \mathbb{P}(X_2=0)=\frac{1}{2}, \quad \mathbb{P}(X_4=0)=\frac{5}{8}.\label{eq:ex1}
\end{equation}
The result is consistent with his claim \cite{gettrick}, i.e., $\mathbb{P}(X_{2t}=0) \ge 1/2$ for any $t \ge 0$. In addition, if $\alpha_3=1,\alpha_1=\alpha_2=\alpha_4=0$, we have
\begin{equation}
 \ket{\psi_4(0)}=\frac{1}{4}\left[\begin{array}{c}
		  2\\-1\\2\\1
		       \end{array}\right].\label{eq:ex2}
\end{equation}
This is equivalent to his computation (2.13) in \cite{gettrick}.

In order to obtain the limit theorems, we introduce the Fourier transform $\ket{\hat{\Psi}_{t}(k)}\,(k\in\left[-\pi,\pi\right))$ of $\ket{\psi_t(x)}$ as follows:
\begin{equation*}
 \ket{\hat{\Psi}_{t}(k)}=\sum_{x\in\mathbb{Z}} e^{-ikx}\ket{\psi_t(x)}.
\end{equation*}
By the inverse Fourier transform, we have
\begin{equation*}
 \ket{\psi_t(x)}=\int_{-\pi}^{\pi}e^{ikx}\ket{\hat\Psi_{t}(k)}\,\frac{dk}{2\pi}.
\end{equation*}
The time evolution of $\ket{\hat{\Psi}_{t}(k)}$ is
\begin{equation}
 \ket{\hat{\Psi}_{t+1}(k)}=\hat U(k)\ket{\hat{\Psi}_{t}(k)},\label{eq:timeevo}
\end{equation}
where $\hat U(k)=R(k)U$ and
\begin{equation*}
 R(k)=\left[\begin{array}{cccc}
       e^{ik}&0&0&0 \\
	     0&e^{ik}&0&0 \\
	     0&0&e^{-ik}&0\\
	     0&0&0&e^{-ik}\\
	    \end{array}\right].
\end{equation*}
From (\ref{eq:timeevo}), we see that
\begin{equation*}
 \ket{\hat{\Psi}_{t}(k)}=\hat U(k)^t\ket{\hat\Psi_{0}(k)}.
\end{equation*}
The probability distribution can be written as
\begin{equation*}
 \mathbb{P}(X_t=x)=\left|\left|\,\int_{-\pi}^{\pi}\hat U(k)^t\ket{\hat\Psi_{0}(k)}e^{ikx}\,\frac{dk}{2\pi}\,\right|\right|^2.
\end{equation*}
In the present paper we take the initial state as
\begin{equation*}
 \ket{\psi_0(x)}=\left\{\begin{array}{ll}
		 \!{}^T[\,\alpha, \,\beta,\,\gamma,\,\delta\,]& (x=0)\\[2mm]
			\!{}^T[\,0,\,0,\,0,\,0\,]& (x\neq 0)
		       \end{array}\right.,
\end{equation*}
where $|\alpha|^2+|\beta|^2+|\gamma|^2+|\delta|^2=1$. We should note that $\ket{\hat\Psi_{0}(k)}=\ket{\psi_0(0)}$.

%\clearpage

%%%%%%%%%%%%%%%%%%%%%%%%%%%%%%%%%%%%%%%%%%%%%%%%%%%%%%%
\section{Limit theorems for the 4-state Hadamard walk}

From now on we focus on the 4-state QW with $a=b=c=-d=1/\sqrt{2}$. That is,
\begin{equation*}
 U=\frac{1}{\sqrt{2}}\left[\begin{array}{cccc}
		      0&0 &1 &1 \\1&-1&0&0 \\ 1&1&0&0 \\ 0&0&1&-1
			   \end{array}\right].
\end{equation*}
In this case, (\ref{eq:coin_2}) becomes
\begin{align*}
 \ket{n,0}&\longrightarrow \frac{1}{\sqrt{2}}(\ket{n,0} + \ket{n,1}),\\
 \ket{n,1}&\longrightarrow \frac{1}{\sqrt{2}}(\ket{n,0} - \ket{n,1}),\\
 \ket{n,2}&\longrightarrow \frac{1}{\sqrt{2}}(\ket{n,2} + \ket{n,3}),\\
 \ket{n,3}&\longrightarrow \frac{1}{\sqrt{2}}(\ket{n,2} - \ket{n,3}),
\end{align*}
which are considered as dynamics of the model corresponding to the 2-state (usual) Hadamard walk. So, this QW is called the 4-state Hadamard walk here.

\begin{figure}[h]
 \begin{center}
 \begin{minipage}{60mm}
  \begin{center}
   \includegraphics[scale=0.4]{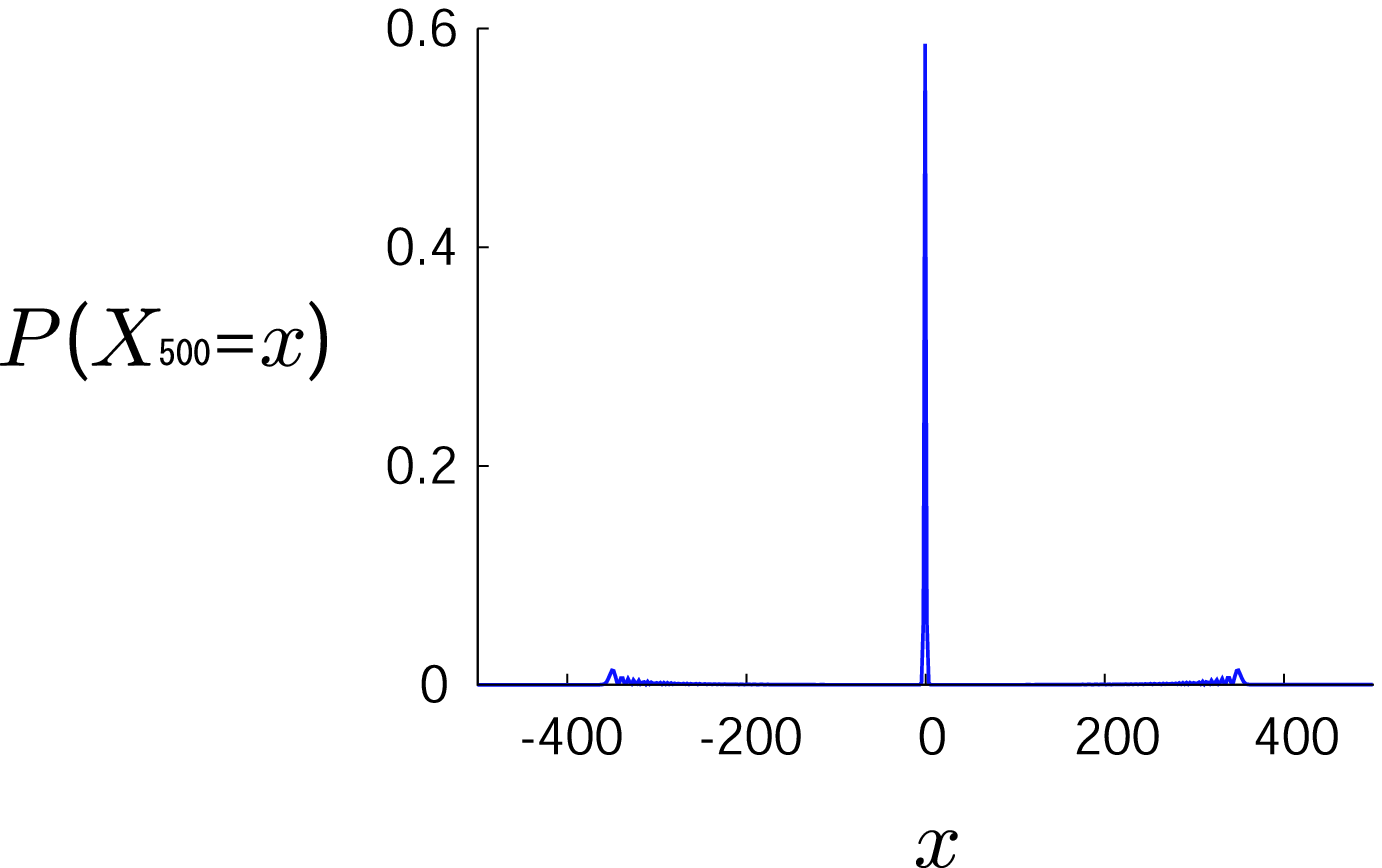}\\
   {(a) $\ket{\psi_0(0)}={}^T[1/2,1/2,1/2,1/2]$}
  \end{center}
 \end{minipage}%\hspace{1cm}
 \begin{minipage}{60mm}
  \begin{center}
   \includegraphics[scale=0.4]{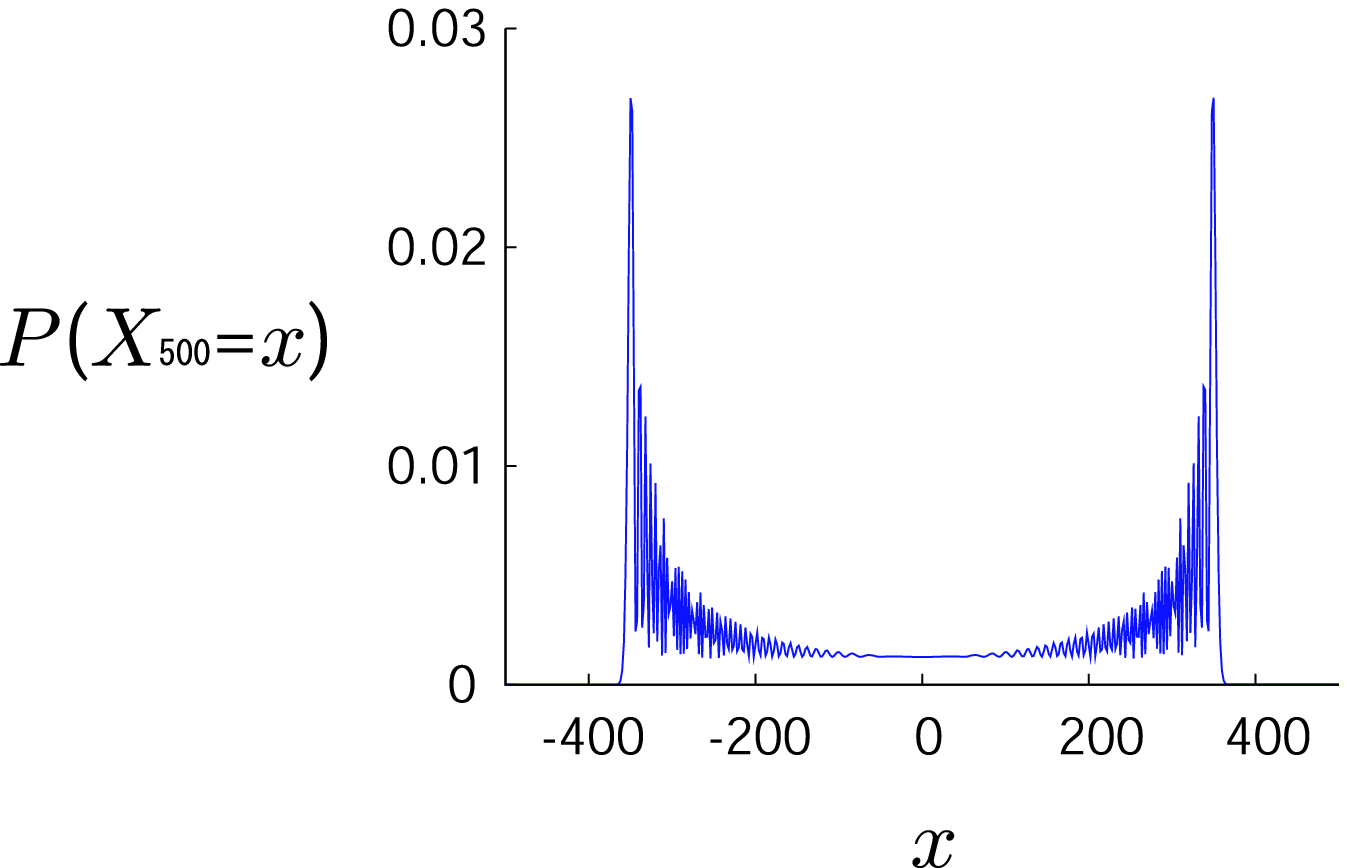}\\
   {(b) $\ket{\psi_0(0)}={}^T[1/2,-1/2,-1/2,1/2]$}
  \end{center}
 \end{minipage}
 \vspace{5mm}
 \fcaption{The probability distributions at time $t=500$.}
  \label{fig:distribution}
 \end{center}
\end{figure}

\begin{figure}[h]
 \begin{center}
 \begin{minipage}{60mm}
  \begin{center}
   \includegraphics[scale=0.5]{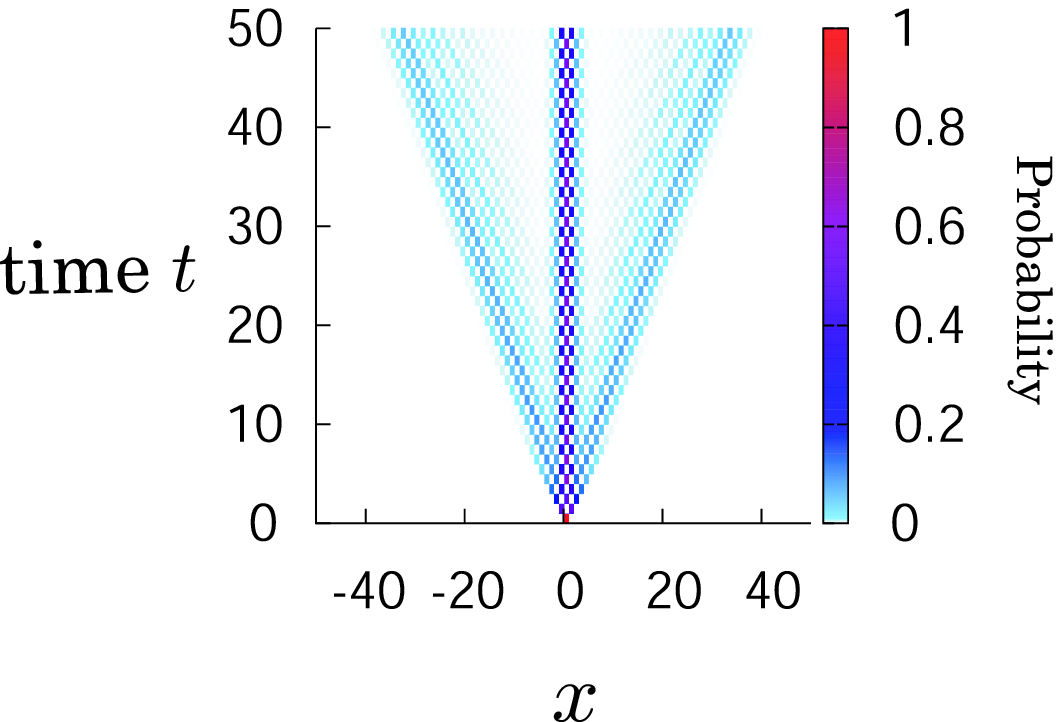}\\
   {(a) $\ket{\psi_0(0)}={}^T[1/2,1/2,1/2,1/2]$}
  \end{center}
 \end{minipage}%\hspace{1cm}
 \begin{minipage}{60mm}
  \begin{center}
   \includegraphics[scale=0.5]{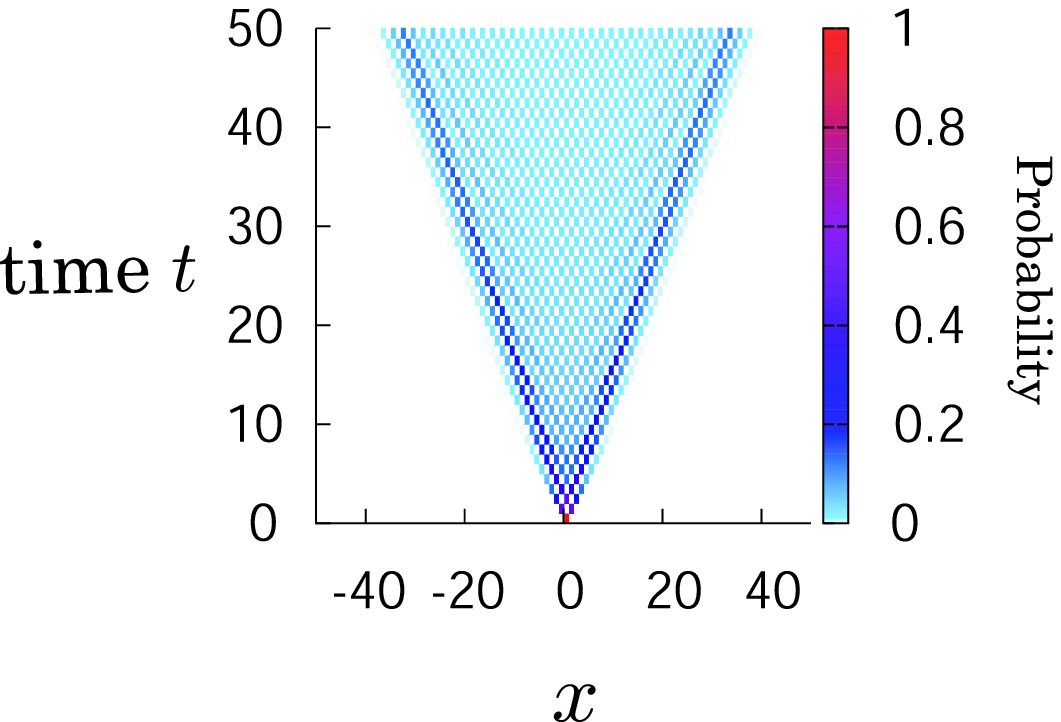}\\
   {(b) $\ket{\psi_0(0)}={}^T[1/2,-1/2,-1/2,1/2]$}
  \end{center}
 \end{minipage}
 \vspace{5mm}
 \fcaption{The behavior of the probability distributions for time $t$ by density plot.}
  \label{fig:dis_time}
 \end{center}
\end{figure}

In this section we present two limit theorems for QW. Both Figures \ref{fig:distribution} and \ref{fig:dis_time} depict the typical probability distributions of the walk. 
\vspace{1cm}

\clearpage

%%%%%%%%%%%%%%%%%%%%%%%%%%%%%%%%%%%THEOREM 1%%%%%%%%%%%%%%%%%%%%%%%%%%%%%%%%%%%%%%%%%%%%%%%%%%%%%%%%%

For the 4-state Hadamard walk, we obtain the following stationary distribution for any initial state.\\

\noindent{\bf Theorem 1:  }
\begin{align*}
 \lim_{t\rightarrow\infty}\mathbb{P}(X_{2t}=0)=&2-\sqrt{2}-(\sqrt{2}-1)(|\beta|^2+|\delta|^2)\nonumber\\
 &+2(\sqrt{2}-1)\Re(\alpha\bar\gamma)+(3-2\sqrt{2})\Re((\alpha-\gamma)(\bar\beta-\bar\delta)),
\end{align*}
 
\begin{equation*}
 \lim_{t\rightarrow\infty}\mathbb{P}(X_{2t}=x)=\left\{\begin{array}{cl}
				       (3-2\sqrt{2})^{|x|-1}K(\alpha,\beta,\gamma,\delta)& (x=2,4,\ldots), \\
					      (3-2\sqrt{2})^{|x|-1}K(\alpha,-\delta,\gamma,-\beta)& (x=-2,-4,\ldots),\\
					      0& (x=\pm 1,\pm3,\ldots),
					     \end{array}\right.
\end{equation*}

\begin{equation*}
 \lim_{t\rightarrow\infty}\mathbb{P}(X_{2t+1}=x)=\left\{\begin{array}{cl}
					 M(\alpha,\beta,\gamma,\delta)& (x=1),\\
						M(\gamma,\delta,\alpha,\beta)& (x=-1),\\
				       (3-2\sqrt{2})^{|x|-1}K(\alpha,\beta,\gamma,\delta)& (x=3,5,\ldots), \\
					      (3-2\sqrt{2})^{|x|-1}K(\alpha,-\delta,\gamma,-\beta)& (x=-3,-5,\ldots),\\
					      0&(x=0,\pm 2,\pm 4,\ldots),
					     \end{array}\right.
\end{equation*}
where $\bar{z}$ is the complex conjugate, $\Re(z)$ is the real part of $z\in\mathbb{C}$ and
\begin{align*}
 M(\alpha,\beta,\gamma,\delta)=&\frac{8-5\sqrt{2}}{2}|\alpha|^2+\frac{2-\sqrt{2}}{2}|\beta|^2
 +(3-2\sqrt{2})|\gamma|^2+(17-12\sqrt{2})|\delta|^2\\
 &+(\sqrt{2}-1)\Re(\alpha\overline{\beta})-\frac{12-9\sqrt{2}}{2}\Re(\alpha\overline{\gamma})-\frac{30-21\sqrt{2}}{2}\Re(\alpha\overline{\delta})\\
 &+\frac{4-3\sqrt{2}}{2}\Re(\beta\,\overline{\gamma})+\frac{10-7\sqrt{2}}{2}\Re(\beta\overline{\delta})+(14-10\sqrt{2})\Re(\gamma\overline{\delta}),\\[5mm]
 K(\alpha,\beta,\gamma,\delta)=&3-2\sqrt{2}+2(\sqrt{2}-1)|\beta|^2+2(7-5\sqrt{2})|\delta|^2\nonumber\\
 &+2\left\{(\sqrt{2}-1)\Re((\alpha-\gamma)\bar\beta)-(7-5\sqrt{2})\Re((\alpha-\gamma)\bar\delta)\right.\nonumber\\
 &\left.+(3-2\sqrt{2})\Re(\beta\bar\delta-\alpha\bar\gamma)\right\}.
\end{align*}

\vspace{1cm}

\begin{figure}[h]
 \begin{center}
 \begin{minipage}{60mm}
  \begin{center}
   \includegraphics[scale=0.4]{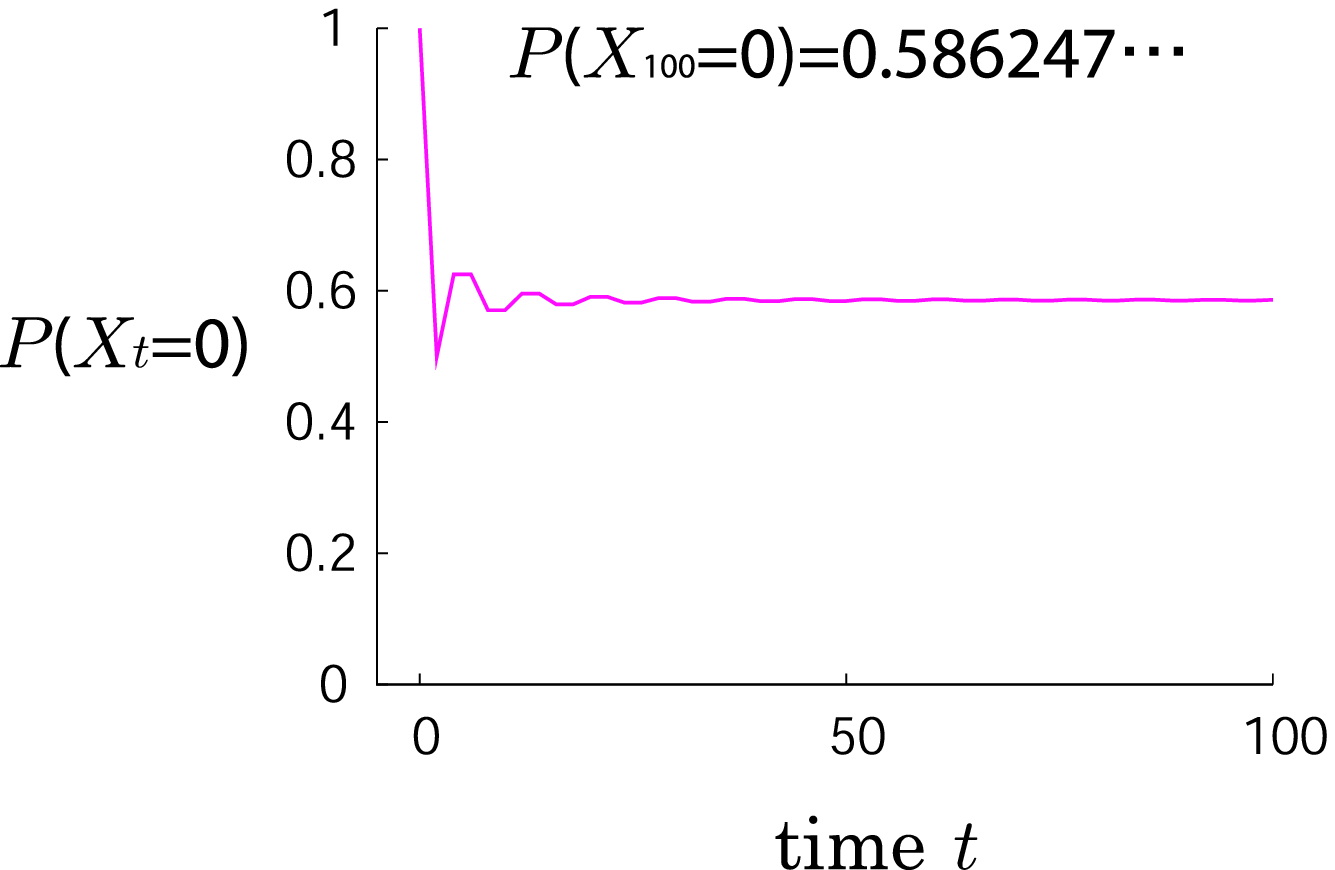}\\
   {(a) $\ket{\psi_0(0)}={}^T[1/2,1/2,1/2,1/2]$}
  \end{center}
 \end{minipage}%\hspace{1cm}
 \begin{minipage}{60mm}
  \begin{center}
   \includegraphics[scale=0.4]{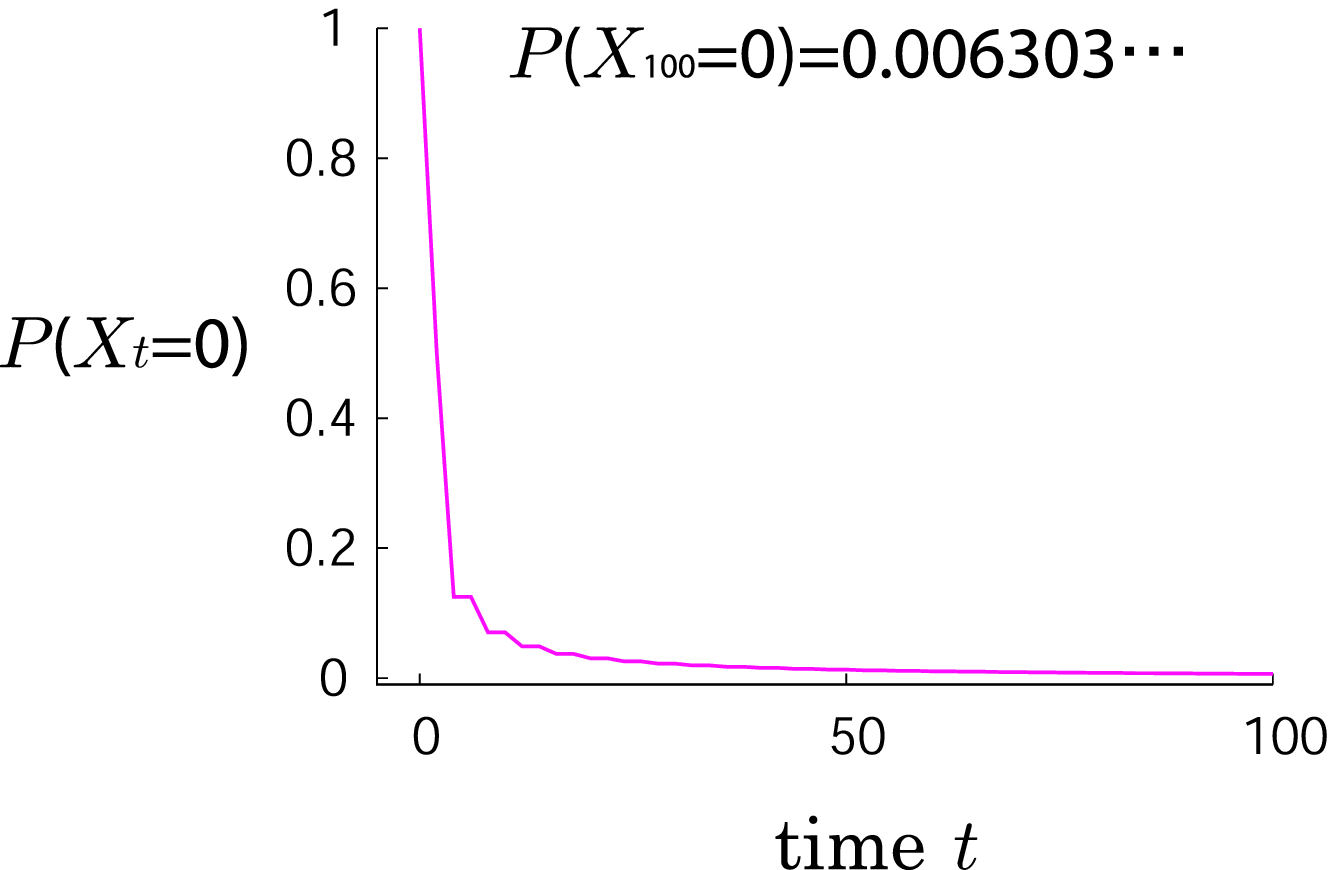}\\
   {(b) $\ket{\psi_0(0)}={}^T[1/2,-1/2,-1/2,1/2]$}
  \end{center}
 \end{minipage}
  \vspace{5mm}
 \fcaption{The behavior of the probability $\mathbb{P}(X_t=0)$ for even time $t$.}
  \label{fig:p0}
 \end{center}
\end{figure}

For $\alpha=\beta=\gamma=\delta=1/2$, Mc Gettrick \cite{gettrick} showed $\mathbb{P}(X_{2t}=0)\geq 1/2$ for any $t \ge 0$ by induction on the time step. Corresponding to his result, Theorem 1 gives $\lim_{t\rightarrow\infty}\mathbb{P}(X_{2t}=0)=2-\sqrt{2}=0.58578\cdots$, see Figure \ref{fig:p0} (a). Then localization occurs. On the other hand, if $\alpha=-\beta=-\gamma=\delta=\pm 1/2$, then Theorem 1 also implies $\lim_{t\rightarrow\infty}\mathbb{P}(X_{2t}=0)=0$, see Figure \ref{fig:p0} (b). Therefore, localization does not occur. This is in sharp contrast to the previous case.

\clearpage

%\clearpage

%%%%%%%%%%%%%%%%%%%%%%%%%%%%%%%%%%THEOREM 2%%%%%%%%%%%%%%%%%%%%%%%%%%%%%%%%%%%%%%%%%%%%%%%%%%%%%%%%%%
The weak limit measure of the rescaled (usual) 2-state Hadamard walk does not have $\delta$-measure corresponding to localization. In fact, the following result was given by Konno \cite{konno_2002_1,konno_2005_1} for any initial state $\ket{\psi_0(0)} = {}^T [\alpha, \beta]$ : for $- \infty < a \le b < \infty,$ 
\begin{equation*}
\lim_{t\rightarrow\infty} \mathbb{P} \left( a \leq \frac{X_t}{t} \leq b \right) =\int_{a}^{b} \left\{ 1 - \left( |\alpha|^2 - |\beta|^2 + \alpha\bar\beta + \bar\alpha\beta \right) x \right\} f_K (x) \,dx,
\end{equation*}
where 
\begin{equation*}
f_K (x) = \frac{1}{\pi(1-x^2)\sqrt{1-2x^2}} \,I_{(-\frac{1}{\sqrt{2}},\frac{1}{\sqrt{2}})}(x),
\end{equation*}
and $I_{A}(x)=1$ if $x\in A$, $I_{A}(x)=0$ if $x\notin A$. In contrast to the above result, our weak limit measure has a $\delta$-measure.\\

\noindent{\bf Theorem 2:  } 
For $- \infty < a \le b < \infty,$ 
\begin{equation*}
 \lim_{t\rightarrow\infty} \mathbb{P} \left( a \leq \frac{X_t}{t} \leq b \right) = \int_{a}^{b} \left\{ \Delta\delta_0(x)+ (c_0+c_1x+c_2x^2) f_K (x) \right\} \,dx,
\end{equation*}
where $\delta_0(x)$ denotes Dirac's $\delta$-function at the origin. Here $\Delta,c_0,c_1,c_2$ are determined by initial state $\ket{\psi_0(0)} = {}^T [\alpha, \beta, \gamma, \delta]$ as follows:
\begin{align*}
 \Delta=&1-\frac{\sqrt{2}}{4}
	   +\frac{1}{2}\left\{(\sqrt{2}-2)(|\beta|^2+|\delta|^2)+(2-\sqrt{2})\Re((\alpha-\gamma)(\bar\beta-\bar\delta))\right.\nonumber\\
 &\qquad\qquad\qquad\left.+\sqrt{2}\Re(\alpha\bar\gamma)-(4-3\sqrt{2})\Re(\beta\bar\delta)\right\},\\
 c_0=&\frac{1}{2}-\Re(\alpha\bar\gamma+\beta\bar\delta),\\
 c_1=&|\delta|^2-|\beta|^2+\Re((\alpha-\gamma)(\bar\beta+\bar\delta)),\\
 c_2=&|\beta|^2+|\gamma|^2-\frac{1}{2}+\Re((\alpha-\gamma)(\bar\delta-\bar\beta)+\alpha\bar\gamma+3\beta\bar\delta).
\end{align*}

\vspace{1cm}

\begin{figure}[h]
 \begin{center}
 \begin{minipage}{60mm}
  \begin{center}
   \includegraphics[scale=0.4]{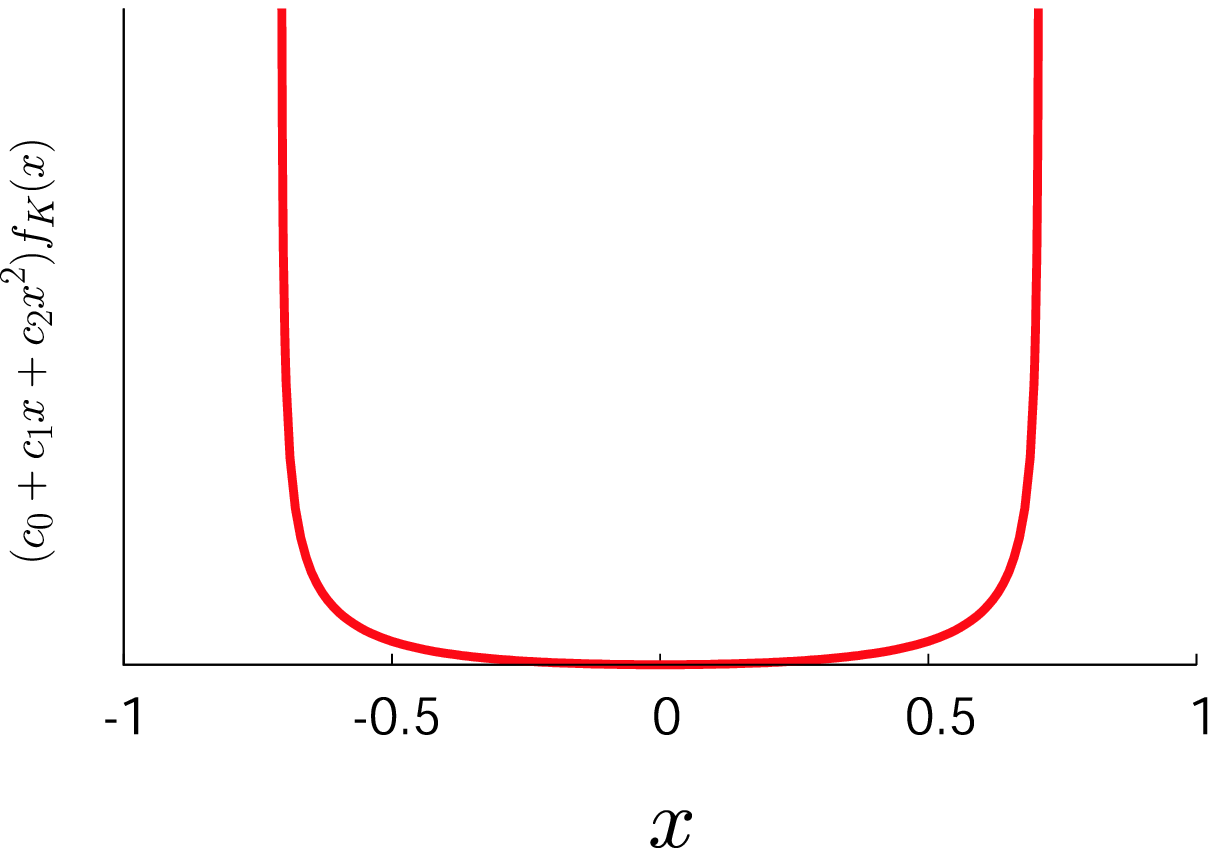}\\
   {(a) $\ket{\psi_0(0)}={}^T[1/2,1/2,1/2,1/2]$}
  \end{center}
 \end{minipage}%\hspace{1cm}
 \begin{minipage}{60mm}
  \begin{center}
   \includegraphics[scale=0.4]{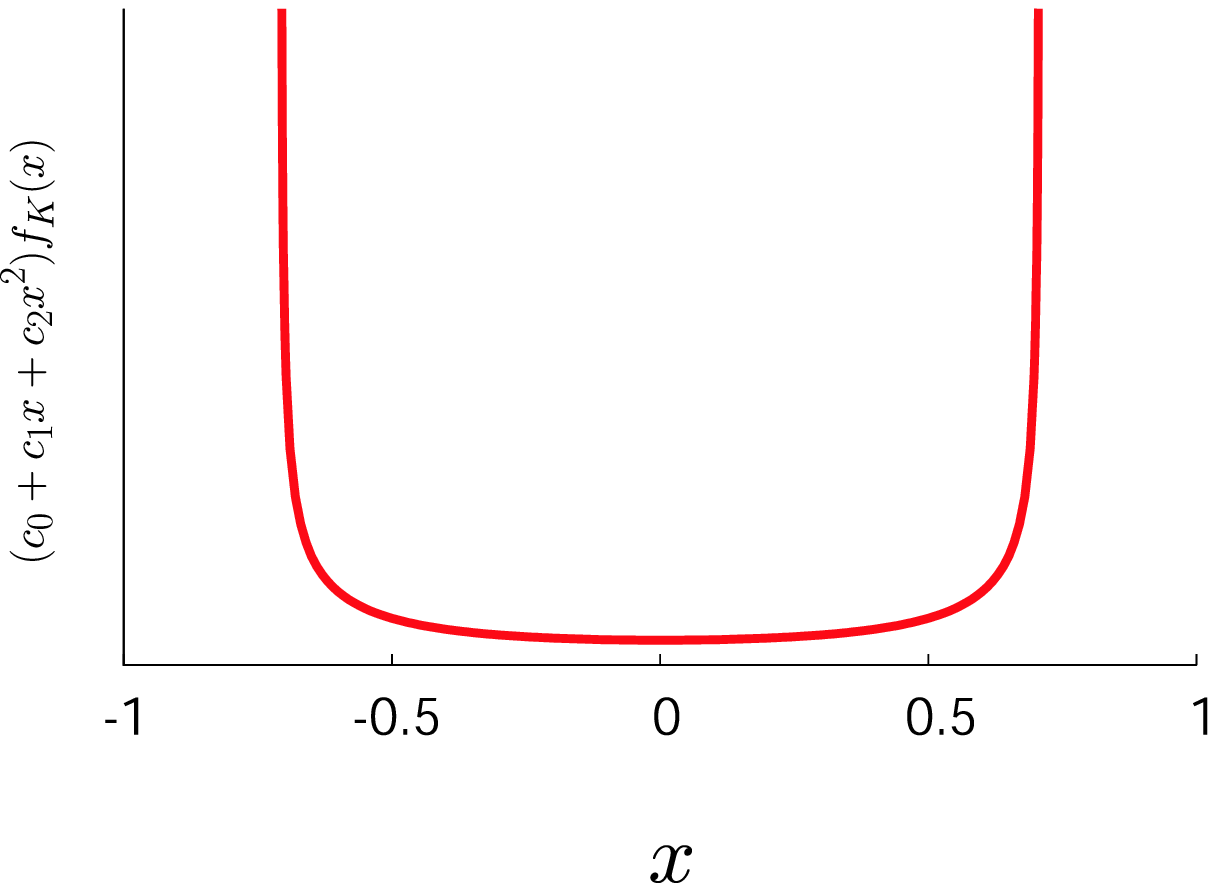}\\
   {(b) $\ket{\psi_0(0)}={}^T[1/2,-1/2,-1/2,1/2]$}
  \end{center}
 \end{minipage}
  \vspace{5mm}
 \fcaption{The limit density function $(c_0+c_1x+c_2x^2)f_K (x)$.}
  \label{fig:density}
 \end{center}
\end{figure}

As for the limit density function $(c_0+c_1x+c_2x^2)f_K (x)$, see Figure \ref{fig:density}. In addition, we have $\Delta=1/\sqrt{2}$ in Figure \ref{fig:density} (a), while we have $\Delta=0$ in Figure \ref{fig:density} (b).

\noindent
As a relation between Theorems 1 and 2, the following equation can be obtained 
\begin{equation*}
 \sum_{x\in\mathbb{Z}}\lim_{t\rightarrow\infty}\mathbb{P}(X_{t}=x)=\Delta.
\end{equation*}
The result for a 3-state QW without memory in Inui et al. \cite{inui1} is similar to Theorem 2. Moreover, localization of a multi-state QW was reported in \cite{inui3}. 
%\clearpage

%%%%%%%%%%%%%%%%%%%%%%%%%%%%%%     PROOF     %%%%%%%%%%%%%%%%%%%%%%%%%%%%%%%%%%%%%%%%
\section{Proofs of theorems}

In this section we will prove Theorems 1 and 2 in Section 3. Our approach is based on the Fourier analysis given by Grimmett et al. \cite{grimmett}.

\subsection{Proof of Theorem 1}

The eigenvalues $\lambda_j(k)\,(j=1,2,3,4)$ of $\hat U(k)$ can be computed as
\begin{equation*}
 \lambda_1(k)=1,\,\lambda_2(k)=-1,\,\lambda_3(k)=\frac{-\cos k+i\sqrt{1+\sin^2 k}}{\sqrt{2}},\,\lambda_4(k)=\frac{-\cos k-i\sqrt{1+\sin^2 k}}{\sqrt{2}}.
\end{equation*}
The eigenvector $\ket{v_j(k)}$ corresponding to $\lambda_j(k)$ is
\begin{equation*}
  \ket{v_j(k)}=\frac{1}{\sqrt{N_j(k)}}\left[\begin{array}{c}
   e^{ik}\left\{\sqrt{2}\lambda_j(k)+e^{ik}\right\}\left\{\lambda_j(k) e^{ik}+\sqrt{2}\right\}\\[3mm]
	 e^{2ik}\left\{\lambda_j(k) e^{ik}+\sqrt{2}\right\}\\[3mm]
	 \lambda_j(k)\left\{\sqrt{2}\lambda_j(k)+e^{ik}\right\}\left\{\sqrt{2}\lambda_j(k) e^{ik}+1\right\}\\[3mm]
	 \lambda_j(k)\left\{\sqrt{2}\lambda_j(k) +e^{ik}\right\}
	\end{array}\right],
\end{equation*}
where $N_j(k)$ is the normalized constant.
The Fourier transform $\ket{\hat\Psi_{0}(k)}$ is expressed by $\ket{v_j (k)}$ as follows:
\begin{equation*}
 \ket{\hat\Psi_{0}(k)}=\sum_{j=1}^4\braket{v_j(k)|\hat\Psi_{0}(k)}\ket{v_j(k)}.
\end{equation*}
Therefore we have
\begin{equation*}
 \ket{\hat\Psi_{t}(k)}=\hat U(k)^t\ket{\hat\Psi_{0}(k)}=\sum_{j=1}^4\lambda_j(k)^t\braket{v_j(k)|\hat\Psi_{0}(k)}\ket{v_j(k)}.
\end{equation*}
By the inverse Fourier transform,
\begin{equation*}
 \ket{\psi_{t}(x)}=\sum_{j=1}^4\int_{-\pi}^{\pi}\lambda_j(k)^t\braket{v_j(k)|\hat\Psi_{0}(k)}\ket{v_j(k)}e^{ikx}\,\frac{dk}{2\pi}.
\end{equation*}
From a similar argument in \cite{inui1}, using the Riemann-Lebesgue lemma, we see
\begin{equation}
 \ket{\psi_{t}(x)}\sim\int_{-\pi}^{\pi}\braket{v_1(k)|\hat\Psi_{0}(k)}\ket{v_1(k)}e^{ikx}\,\frac{dk}{2\pi}+(-1)^t\int_{-\pi}^{\pi}\braket{v_2(k)|\hat\Psi_{0}(k)}\ket{v_2(k)}e^{ikx}\,\frac{dk}{2\pi},\label{eq:psi_t_sim}
\end{equation}
where $g(t)\sim h(t)$ denotes $\lim_{t\rightarrow\infty}g(t)/h(t)=1$.
By (\ref{eq:psi_t_sim}), we obtain
\begin{align}
 \ket{\psi_{t}(0)}\,\sim&\,\frac{1+(-1)^t}{8}
  \left[\begin{array}{c}
   (4-\sqrt{2})\alpha+(2-\sqrt{2})\beta+\sqrt{2}\gamma-(2-\sqrt{2})\delta\\
	 (2-\sqrt{2})\alpha+\sqrt{2}\beta-(2-\sqrt{2})\gamma-(4-3\sqrt{2})\delta\\
	 \sqrt{2}\alpha-(2-\sqrt{2})\beta+(4-\sqrt{2})\gamma+(2-\sqrt{2})\delta\\
	 -(2-\sqrt{2})\alpha-(4-3\sqrt{2})\beta+(2-\sqrt{2})\gamma+\sqrt{2}\delta
	\end{array}\right],\label{eq:psi_0_sim}\\
 \ket{\psi_{t}(1)}\,\sim&\,\frac{1-(-1)^t}{8}
  \left[\begin{array}{c}
   (2-\sqrt{2})\alpha+\sqrt{2}\beta-(2-\sqrt{2})\gamma-(4-3\sqrt{2})\delta\\
	 (4-3\sqrt{2})\alpha-(2-\sqrt{2})\beta-(4-3\sqrt{2})\gamma-(10-7\sqrt{2})\delta\\
	 -(2-3\sqrt{2})\alpha+\sqrt{2}\beta+(2-\sqrt{2})\gamma+(4-3\sqrt{2})\delta\\
	 \sqrt{2}\alpha-(2-\sqrt{2})\beta+\sqrt{2}\gamma-(2-\sqrt{2})\delta
	\end{array}\right],\label{eq:psi_1_sim}\\
 \ket{\psi_{t}(-1)}\,\sim&\,\frac{1-(-1)^t}{8}
  \left[\begin{array}{c}
   (2-\sqrt{2})\alpha+(4-3\sqrt{2})\beta-(2-3\sqrt{2})\gamma+\sqrt{2}\delta\\
	 \sqrt{2}\alpha-(2-\sqrt{2})\beta+\sqrt{2}\gamma-(2-\sqrt{2})\delta\\
	 -(2-\sqrt{2})\alpha-(4-3\sqrt{2})\beta+(2-\sqrt{2})\gamma+\sqrt{2}\delta\\
	 -(4-3\sqrt{2})\alpha-(10-7\sqrt{2})\beta+(4-3\sqrt{2})\gamma-(2-\sqrt{2})\delta
	\end{array}\right],\label{eq:psi_-1_sim}
\end{align}
and for $x=2,3,\ldots$,
\begin{align}
 \ket{\psi_{t}(x)}\,\sim&\,\frac{(\sqrt{2}-1)^x}{4\sqrt{2}(3-2\sqrt{2})}\left\{(\sqrt{2}-1)\alpha+\beta-(\sqrt{2}-1)\gamma+(3-2\sqrt{2})\delta\right\}\nonumber\\
 &\qquad\times\left\{(-1)^x+(-1)^t\right\}
  \left[\begin{array}{c}
   1-\sqrt{2}\\ 3-2\sqrt{2}\\ \sqrt{2}-1\\ 1
	\end{array}\right],\label{eq:psi_x+_sim}
\end{align}
for $x=-2,-3,\ldots$,
\begin{align}
 \ket{\psi_{t}(x)}\,\sim&\,-\frac{(\sqrt{2}-1)^{-x}}{4\sqrt{2}(3-2\sqrt{2})}\left\{(\sqrt{2}-1)\alpha-(3-2\sqrt{2})\beta-(\sqrt{2}-1)\gamma-\delta\right\}\nonumber\\
 &\qquad\times\left\{(-1)^x+(-1)^t\right\}
  \left[\begin{array}{c}
   \sqrt{2}-1\\ 1\\ 1-\sqrt{2}\\ 3-2\sqrt{2}
	\end{array}\right].\label{eq:psi_x-_sim}
\end{align}
Combining (\ref{eq:psi_0_sim}), (\ref{eq:psi_1_sim}), (\ref{eq:psi_-1_sim}), (\ref{eq:psi_x+_sim}) with (\ref{eq:psi_x-_sim}) completes the proof.
\begin{flushright}
%$\Box$ 
\end{flushright}

%%%%%%%%%%%%%%%%%%%%%%%%%%%%%%%%%%%%%%%%%%%%%%%%%%%%%%%%%%%%%%%%%%%
\subsection{Proof of Theorem 2}

%We calculate the characteristic function $E(e^{i \xi X_{t}/t})$, 
%where $\xi$ is a real number. 

First the {\it r}-th moment of $X_t$ becomes
\begin{align*}
 E((X_{t})^r)=&\sum_{x\in \mathbb{Z}}x^r \mathbb{P}(X_{t}=x)\nonumber\\
=&\int_{-\pi}^{\pi}\bra{\hat\Psi_{t}(k)}\left(D^r\ket{\hat\Psi_{t}(k)}\right)\,\frac{dk}{2\pi}\nonumber\\
 =&\int_{-\pi}^{\pi}\sum_{j=1}^4(t)_r\lambda_j(k)^{-r}(D\lambda_j(k))^r\left|\braket{v_j(k)|\hat\Psi_{0}(k)}\right|^2\,\frac{dk}{2\pi}+O(t^{r-1}),
\end{align*}
where $D=i(d/dk)$ and $(t)_r=t(t-1)\times\cdots\times(t-r+1)$.
Let $h_j(k)=D\lambda_j(k)/\lambda_j(k)$. Then we obtain
\begin{align*}
 \lim_{t\rightarrow\infty}E((X_{t}/t)^r)
  =&\int_{-\pi}^{\pi}\sum_{j=1}^4 h_j^r(k) \left| \braket{v_j(k)|\hat\Psi_0(k)} \right|^2\,\frac{dk}{2\pi}\nonumber\\
  =&0^r\Delta + \int_{-\pi}^{\pi}\sum_{j=3}^4 h_j^r(k) \left| \braket{v_j(k)|\hat\Psi_0(k)} \right|^2\,\frac{dk}{2\pi},
\end{align*}
where
\begin{align*}
 \Delta=&\int_{-\pi}^{\pi}\sum_{j=1}^2 \left| \braket{v_j(k)|\hat\Psi_0(k)} \right|^2\,\frac{dk}{2\pi}\nonumber\\
 =&1-\frac{\sqrt{2}}{4}
	   +\frac{1}{2}\left\{(\sqrt{2}-2)(|\beta|^2+|\delta|^2)+(2-\sqrt{2})\Re((\alpha-\gamma)(\bar\beta-\bar\delta))\right.\nonumber\\
 &\qquad\qquad\qquad\left.+\sqrt{2}\Re(\alpha\bar\gamma)-(4-3\sqrt{2})\Re(\beta\bar\delta)\right\}.
\end{align*}
Therefore we get
\begin{align}
 \lim_{t\rightarrow\infty}E((X_{t}/t)^r)=&0^r\Delta+\int_{-\infty}^{\infty} x^r\frac{c_0+c_1x+c_2x^2}{\pi(1-x^2)\sqrt{1-2x^2}}\,I_{(-\frac{1}{\sqrt{2}},\frac{1}{\sqrt{2}})}(x)\,dx\nonumber\\
=&\int_{-\infty}^{\infty} x^r\left\{\Delta\delta_0(x)+\frac{c_0+c_1x+c_2x^2}{\pi(1-x^2)\sqrt{1-2x^2}}\,I_{(-\frac{1}{\sqrt{2}},\frac{1}{\sqrt{2}})}(x)\right\}\,dx
%\nonumber\\
%=&\int_{-\infty}^{\infty} x^r f(x)\,dx
,\label{eq:r-th_mom}
\end{align}
where
\begin{align*}
 c_0=&\frac{1}{2}-\Re(\alpha\bar\gamma+\beta\bar\delta),\\
 c_1=&|\delta|^2-|\beta|^2+\Re((\alpha-\gamma)(\bar\beta+\bar\delta)),\\
 c_2=&|\beta|^2+|\gamma|^2-\frac{1}{2}+\Re((\alpha-\gamma)(\bar\delta-\bar\beta)+\alpha\bar\gamma+3\beta\bar\delta).
\end{align*}
%By (\ref{eq:r-th_mom}), we can compute the characteristic function 
%$E(e^{i \xi X_{t}/t})$ as $t\rightarrow\infty$. 
Thus the proof of Theorem 2 is completed.
\begin{flushright}
%$\Box$ 
\end{flushright}

%%%%%%%%%%%%%%%%%%%%%%%%%%%%%%%%%%%%%%%%%%%%%%%%%%%%%%%
\section{Summary}
In the final section we conclude and discuss our results of the 4-state QW. Mc Gettrick \cite{gettrick} introduced and investigated a new kind of 2-state QWs with one-step memory. We showed that his walk becomes a 4-state QW by relabeling his notation $\ket{n_2,n_1,p}$ (e.g., $\ket{n_2,n_1,p}\rightarrow \ket{n_1,n_1 - n_2 + 1 + p}$). Similarly, any extended version of his walk with $r$-step memory can be considered as a $2^{r+1}$-state QW without memory. In this paper, we obtained two limit theorems for the 4-state Hadamard walk corresponding to the case (c) studied in his paper. From Theorem 1, we found that localization occurs for an initial state. Moreover Theorem 2 implies that $X_t/t$ converges weakly to a random variable with a $\delta$-measure as $t\rightarrow\infty$ for any initial state. One of the interesting future problems is to obtain the limit theorems of the QW for a general $a,b,c,d \in \mathbb{C}$ of $U$ and $m$-state. For an $m$-state QW different from our model, Segawa and Konno \cite{Segawa} presented a convergence theorem for a suitable limit of $t$ and $m\rightarrow\infty$. So we think that it is important to clarify the relation between them.

%%%%%%%%%%%%%%%%%%%%%%%%%%%%%%%%%%%%%%%%%%%%%%%%%%%
%\vspace{5mm}
\nonumsection{Acknowledgements}
\noindent
This work was partially supported by the Grant-in-Aid for Scientific Research (C) of Japan Society for
the Promotion of Science (Grant No. 21540118).

%\clearpage

\nonumsection{References}
\bibliography{main}

\end{document}